\documentclass{article}
\usepackage{spconf,amsmath,amssymb,graphicx}
\usepackage{cite}
\usepackage{mathrsfs}
\usepackage[lined]{algorithm2e}
\usepackage{wasysym}
\usepackage{color}
\usepackage{subfigure}
\usepackage{multirow}
\usepackage{amsthm}
\usepackage{amsopn}
\usepackage{url}
\usepackage{helvet}
\usepackage{bm}
\usepackage{rotating}
\usepackage[utf8]{inputenc}
\usepackage{pgfplots}

\newcommand{\RR}{\mathbb{R}}

\newcommand{\CC}{\mathbb{C}}
\newcommand{\vx}{\mathbf{x}}
\newcommand{\vF}{\mathbf{F}}
\newcommand{\vW}{\mathbf{W}}
\newcommand{\vZ}{\mathbf{Z}}
\newcommand{\vX}{\mathbf{X}}
\newcommand{\vY}{\mathbf{Y}}

\title{Multi-layered Cepstrum for Instantaneous Frequency Estimation}
\name{Chin-Yun Yu and Li Su}
\address{Institute of Information Science, Academia Sinica, Taiwan}

\begin{document}
\maketitle

\begin{abstract}
We propose the multi-layered cepstrum (MLC) method to estimate multiple fundamental frequencies (MF0) of a signal under challenging contamination such as high-pass filter noise. Taking the operation of cepstrum (i.e., Fourier transform, filtering, and nonlinear activation) recursively, MLC is shown as an efficient method to enhance MF0 saliency in a step-by-step manner. Evaluation on a real-world polyphonic music dataset under both normal and low-fidelity conditions demonstrates the potential of MLC. 
\end{abstract}

\section{Introduction}
\label{sec: intro}
Multiple fundamental frequency (MF0) estimation is an essential problem in analyzing various multi-component signals including 
multi-talker speech \cite{pirker2011pitch, wu2003multipitch}, polyphonic music \cite{benetos2013automatic}, 
fetal electrocardiogram (FECG), photoplethysmographic (PPG) signals \cite{zhang2015troika}, and IoT sensor data \cite{li2015eperiodicity}, to name but a few. 
Numerous methods have been proposed to enhance the saliency of F0s in the signal representation, such as the autocorrelation function (ACF) \cite{rabiner1977use}, generalized cepstrum \cite{kobayashi1984spectral}, nonnegative matrix factorization (NMF), MUltiple SIgnal Classification (MUSIC) \cite{christensen2008multi}, 
combined frequency and periodicity (CFP) \cite{su2015combining, su2016exploiting, lin2017wave}, and deep learning approaches \cite{Verma2016FrequencyEF}.
The generalized cepstrum, being one of the earliest proposed MF0 estimation methods \cite{kobayashi1984spectral}, still gains attention in recent years, by showing its effectiveness in suppressing the unwanted harmonics and in localizing the F0 components by combining it with a spectrum \cite{su2015combining, su2016exploiting, peeters2006music}. A recent study further proved that a cepstrum can perfectly localize the instantaneous F0s of a signal with the intrinsic mode function (IMF) type \cite{lin2017wave}, a general type of non-stationary oscillatory signals.

One issue which is relatively less discussed in MF0 estimation is the so-called \emph{convolutional} noise in signals \cite{hermansky1993recognition}. 
Convolutional noise is ubiquitous; it occurs when signals are transmitted through an intermediate channel or device. 
An MF0 algorithm usually fails in the presence of convolutional noise, for example, in the processing of audio signals played by a smartphone, as the response of a smartphone speaker typically behaves like a high-pass filter whose cutoff frequency is around 500 Hz \cite{mauch2013audio,su2015combining}. Such a high-pass filter suppresses the true F0 peaks, as similar to the case of the ``missing fundamental'' effects in low-pitched musical signals \cite{rossing2002science}. In such cases, to identify the true F0 peaks merely from the interleaved harmonics of multiple components is challenging. 

In this paper, we generalize the notion of cepstrum to resolve the above-mentioned issue in MF0 estimation.
The proposed \emph{multi-layered cepstrum} (MLC) refines the saliency of the F0s in a recursive manner. 
In a nutshell, the MLC takes the operation of cepstrum (i.e., Fourier transform, filtering and nonlinear activation) repeatedly: for a time-domain input signal $\mathbf{x} \in \mathbb{R}^N$, $l\geq 0$, the $l^{\textrm{th}}$-layer output of MLC is:
\begin{equation}
\mathbf{z}^{(0)}=\sigma^{(0)}\left(|\vF\vx|\right)\,,\quad \mathbf{z}^{(l)}=\sigma^{(l)}\left(\vW^{(l)}\vF\mathbf{z}^{(l-1)}\right)\,, \nonumber
\end{equation}
where $\vF\in\CC^{N\times N}$ is the $N$-point discrete Fourier transform (DFT) matrix, $\vW\in\RR^{N\times N}$ is a high-pass filter, and $\sigma(\cdot)$ is an element-wise power-scaled rectification unit (see Section \ref{sec:MLC}). 
In the following sections, we will explain how the idea of MLC is motivated, how it works in MF0 estimation, and its structural resemblance to multi-layered perceptrons (MLP). We will then focus on combining two MLCs lying in both the frequency and time domains to achieve MF0 estimation by employing the CFP approach. 
Experiments on both synthetic and real data demonstrates the effectiveness of MLC in MF0 estimation in both normal and contaminated conditions,
and indicate positively the fact that MLC performs better than other methods based on a `shallow' DFT. 
\vspace{-0.3cm}
\section{Motivation of MLC}

The \emph{cepstrum} $\hat{\mathbf{z}}:=\vF^{-1}\sigma(\vF\vx)$ of $\vx$ 
\cite{bogert1963quefrency, Oppenheim_Schafer:2009} is a classic method to estimate the F0 of $\vx$.
A logarithm cepstrum employing $\sigma(x):=\log x$ is commonly used in single F0 estimation, while the \emph{generalized cepstrum} (GC) employing $\sigma(x):=x^{\gamma}$, $0<\gamma\leq 2$, is found more applicable in MF0 estimation, probably because it is numerically more stable than using the logarithm function and is more robust to noise 
\cite{tolonen2000computationally, klapuri2008multipitch}.
Cepstrum-based MF0 estimation is based on an assumption that the fast-varying components in a spectrum (e.g., harmonic sequences) are relevant to F0s, while the slow-varying ones  (e.g., spectral envelope) are irrelevant to F0s. Saliency of F0s can be enhanced in the \emph{quefrency}\footnote{Since the quefrency has the same unit as time, in this paper, the terms quefrency and time are used interchangeably.} domain by taking another Fourier transform and a high-pass filter with a cutoff quefrency at $q_c$ on a nonlinearly-stretched spectrum, where the nonlinear operation is to fit the magnitude of the spectrum to humans' perception scale. 

To motivate the MLC, we revise the above-mentioned assumption to that 1) the fast-varying and \emph{periodic} components in a spectrum are relevant to F0s while others are not, and that 2) such periodic components still remain periodic (and therefore fast-varying) after a Fourier transform, which is a basic property of DFT. These assumptions allow us to consider  
repeating the operation of cepstrum on $\vx$ to iteratively discard the slow-varying or aperiodic components and to purify the F0-relevant components, which should remain after a number of filtering processes. 
This idea has been preliminarily verified in \cite{su2017HSP_DNN}, where the \emph{generalized cepstrum of spectrum} (GCoS), being the $2^{\text{nd}}$layer output of the MLC, behaves as a more succinct feature indicating the F0 peaks than a spectrum. This paper is an extension of \cite{su2017HSP_DNN} 
to a multi-layered time-frequency analysis tool.


Admittedly, the above-mentioned idea is also motivated by deep learning, which catches wide attention recently in MF0 estimation \cite{sigtia2016end, Verma2016FrequencyEF, BittnerDeepSalience17,thickstun2016learning,elowsson2018polyphonic}. Not only achieving state-of-the-art results, deep learning also provides inspiring observation when solving the MF0 estimation problem. For example, results in \cite{Verma2016FrequencyEF} show that the learned networks for melody tracking resemble traditional pitch detection methods: the first layer behaves like a spectral analyzer, while the second layer behaves like a comb filter. It is again intriguing to consider a multi-layered network for MF0 estimation, where the well-known pitch detection blocks such as the Fourier transform and filters are taken directly as subnetworks. 




\section{The MLC-CFP method}


\label{sec:MLC}
We formulate the MLC of a non-stationary signal $\vx$ in the time-frequency or time-quefrency domain. Given 
a window function $\mathbf{h}\in\RR^N$ and hop size $H$, the short-time Fourier transform (STFT) of $\vx$, $\vX:=\vX[k,n]$, is represented as  
$\vX:=\vX[k,n]=\sum^{N-1}_{m=0} \vx[m+nH]\mathbf{h}[m]e^{-\frac{j2\pi km}{N}}$, 
where $\vX\in\CC^{N\times M}$, $M$ is the number of frames, $k$ is the frequency index, and $n$ is the time index.
Denote $|\vX|$ the element-wise absolute value of $\vX$. The $l^{\textrm{th}}$-layer short-time MLC is: 
\begin{equation}
\vZ^{(0)}:=\sigma^{(0)}\left(\left|\vX\right|\right)\,,\quad \vZ^{(l)}:=\sigma^{(l)}\left(\vW^{(l)}\vF\vZ^{(l-1)}\right)\,, \label{eq:klayer_ceps}
\end{equation}
where $\vZ^{(l)}\in\RR^{N\times M}$, $\vZ^{(l)}:=\vZ^{(l)}[k,n]$ for even $l$, $\vZ^{(l)}:=\vZ^{(l)}[q,n]$ for odd $l$, and $q$ is the quefrency index. $\sigma^{(l)}$ is a nonlinear, element-wise power function such that for $\gamma_l>0$, $\sigma^{(l)}\left(x\right) = x^{\gamma_l}$ for $x>0$, and $\sigma^{(l)}\left(x\right) = 0$ for $x\leq 0$. With the $N$-point DFT matrix $\vF$, the frequency resolution of $\vZ^{(l)}[k,n]$ is $f_s/N$ per bin, and the time resolution of $\vZ^{(l)}[q,n]$ is then $1/f_s$ per bin. 
$\mathbf{W}^{(i)}\in\mathbb{R}^{N\times N}$ represents a high-pass filter; it filters out the elements lying below a cutoff frequency (quefrency) $i_c$. More precisely, $\mathbf{W}^{(l)}$ is a diagonal matrix such that $\mathbf{W}^{(l)}[i,i]=1$ when $i>i_c$ and $i<N-i_c$, while $\mathbf{W}^{(l)}[i,i]=0$ otherwise.
For even $l$, $i_c:=k_c$ defines the cutoff frequency at $f_c=k_cf_s/N$, and for odd $l$, $i_c:=n_c$ defines the cutoff quefrency at $q_c=n_c/f_s$. Also note that since $\vZ^{(l)}[:,n]$ is even for $l>2$, we use $\vF$ instead of $\vF^{-1}$ when transforming from the frequency domain to the time domain. 
Many of the basic pitch detection functions are essentially special cases of (\ref{eq:klayer_ceps}). For example, $\vZ^{(0)}$ is the STFT of $\mathbf{x}$, and $\vZ^{(1)}$ is known as the autocorrelation function (ACF) for $\gamma_0=2$ and $\gamma_1=1$. 
(\ref{eq:klayer_ceps}) also resembles a DNN structure with zero bias vector: $\vW^{(l)}\vF$ resembles the fully-connected layer and $\sigma^{(l)}$ resembles the activation function.


To enhance the true F0 components and suppress the high-order harmonics, consider the last two output layers, $\vZ^{(l_e)}$ and $\vZ^{(l_o)}$, of an MLC, where $l_e$ is even and $l_o$ is odd. $\vZ^{(l_e)}$ is in the frequency domain and $\vZ^{(l_o)}$ is in the quefrency domain. Based on the CFP approach, we utilize the duality between them to enhance the saliency of F0 by employing a nonlinear mapping to $\vZ^{(l_o)}$ from the quefrency domain to frequency domain, and then multiplying it by $\vZ^{(l_e)}$ \cite{peeters2006music, su2015combining,su2016exploiting,lin2017wave}: 



\begin{equation}
\vY^{(l_e, l_o)}[k,n]=\vZ^{(l_e)}[k,n]\vZ^{(l_o)}\left[\lfloor N/k\rceil,n\right]
\label{eq:fusion}
\end{equation}
where the time-quefrency representation $\vZ^{(l_o)}[k,n]$ is nonlinearly mapped into the frequency domain, and $\lceil\cdot\rfloor$ is the rounding function. Since $\vZ^{(l_o)}$ suppresses the harmonic peaks in $\vZ^{(l_e)}$ while $\vZ^{(l_e)}$ suppresses the sub-harmonic peaks in $\vZ^{(l_o)}$, the resulting CFP representation $\vY^{(l_e, l_o)}:=\vY^{(l_e, l_o)}[k,n]$ has only true F0 peaks enhanced. The nonlinear mapping of $\vZ^{(l_o)}$ in (\ref{eq:fusion}) can be implemented by a filterbank following a custom-defined indexes of $k$; see \cite{su2016exploiting,su2017HSP_DNN} for details. We name the above process as the MLC-CFP algorithm hereafter.

\begin{figure*}[t]
\includegraphics[width=\textwidth]{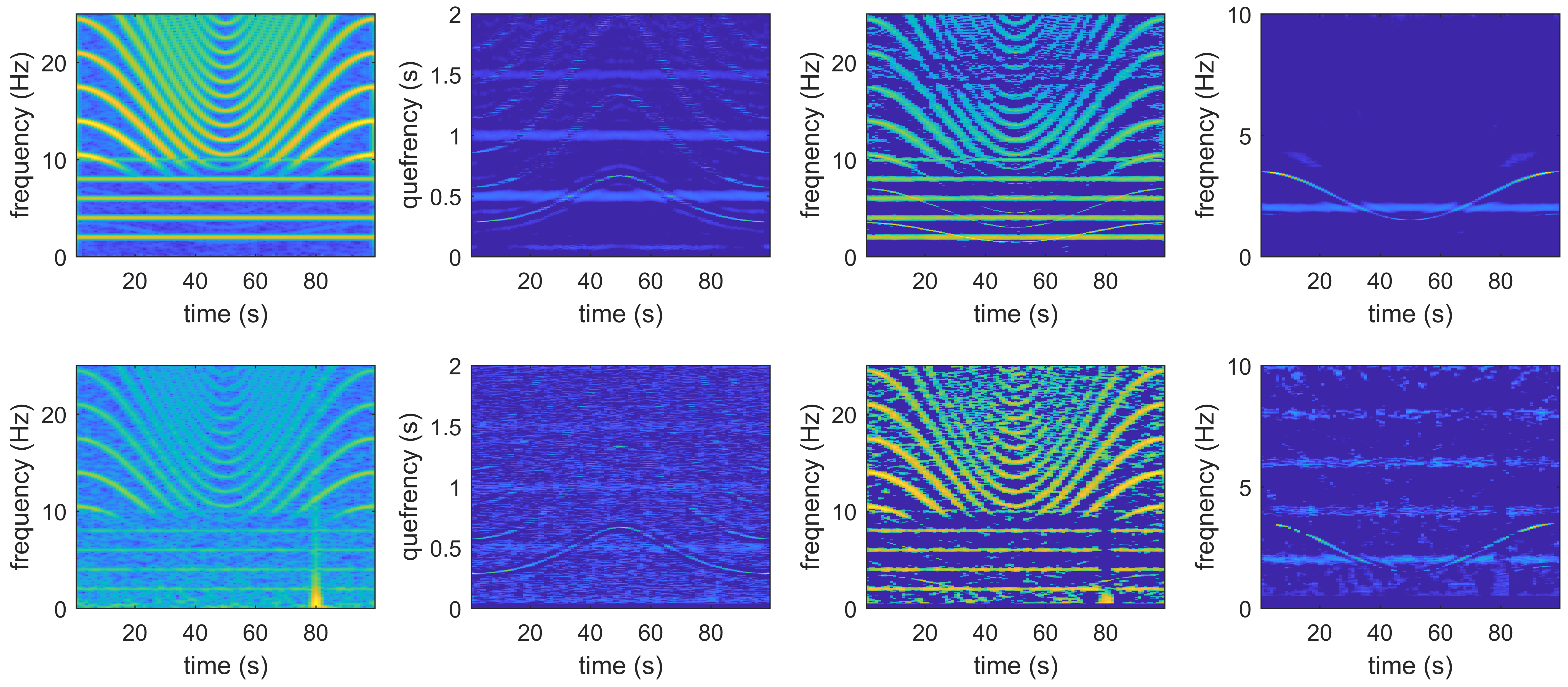}
\centering
\caption{Simulation result. Upper: $\vx$; lower: $\hat{\vx}$. From left to right: $\vZ^{(0)}$ (power-scale spectrogram), $\vZ^{(1)}$ (GC), $\vZ^{(2)}$ (GCoS), and $\vY^{(2,1)}$ (CFP representation). The window size is 8 seconds and the hop size is 1 second. }
\label{fig:2layer_ddft}
\end{figure*}

\section{Numerical results}

We illustrate the effectiveness of the MLC-CFP method in emphasizing F0 peaks of a simulated signal contaminated with both additive and convolutional noise. Given two components, $\vx_1$ and $\vx_2$, both are sampled at $f_s=1$ kHz. $\vx_1$ is a square wave having 20\% duty cycle, and has a constant F0 at $2$ Hz. $\vx_2$ is a frequency-modulated sawtooth wave with positive ramp and vertical drop, and has time-varying F0 represented as $f_0[n] = 2.5+\cos(2\pi n/10f_s)$. To simulate the effect of convolutional noise, $\vx_1$ is passed through a 10th-order Butterworth low-pass filter, and $\vx_2$ is passed through a 10th-order Butterworth high-pass filter. Both filters have a cutoff frequency at 10 Hz. We consider the 2-layer MLC of two sample signals: the first one is $\vx = \vx_1 + \vx_2$, which is with convolutional noise only, while the second one, $\tilde{\vx}=\vx_1+\vx_2+\mathbf{n}+\mathbf{d}$, is further contaminated with additive pink noise and impulse noise, where $\mathbf{n}$ is additive pink noise making the SNR of $\tilde{\vx}$ be 10 dB and $\mathbf{d}$ is an impulse at the 80th second.


Fig. \ref{fig:2layer_ddft} shows the STFT ($\vZ^{(0)}$), GC ($\vZ^{(1)}$), GCoS ($\vZ^{(2)}$), and CFP representation ($\vY^{(2,1)}$) of $\vx$ and $\hat{\vx}$, respectively. The parameters $[\gamma_0, \gamma_1, \gamma_2] = [0.24, 0.6, 1]$ are used. For the STFT of both samples, the F0 peaks and the low-order harmonics of $\vx_2$ are all unseen, and only the first 5 multiples of $\vx_1$ are visible due to low-pass filtering. To capture the true F0 of $\vx_2$ from the STFT is very challenging. However, such information can be found in the GC: although high-pass filtering noise attenuates the true F0 component in the spectrum, it does not change the period of the signal. Such a phenomenon can also be seen in the case with pink noise. The true F0 trajectory then appears in the GCoS with high resolution; this is because the nonlinear scaling of the GC provides the mechanism to \emph{down-mixing} the frequency when it is transformed back to the frequency domain. Interestingly, we can see that although the true F0 of $\vx_2$ is missed, the resulting F0 trajectory of $\vx_2$ has resolution even higher than the one of $\vx_1$. Since most of the harmonic peaks of $\vx_1$ are eliminated, its spectrum cannot support better resolution when transformed to the quefrency domain. Finally, the CFP representation computed through (\ref{eq:fusion}) clearly indicates the true F0 trajectories of $\vx_1$ and $\vx_2$ with less interference.

\section{Real-data Experiments}

\begin{table*}[t]
\centering
\caption{Comparison of different layer combinations and testing method on the Bach10 dataset.}
\label{table:comparison}
\small
\begin{tabular}{|c|ccccccccc|cccc|ccc|}
\hline
Layers              & \multicolumn{9}{c|}{Brute Force}                                   & \multicolumn{4}{c|}{Greedy}
& \multicolumn{3}{c|}{SGD (10-fold CV)}\\ \cline{2-17}
$\{L, L+1\}$ & $\gamma_0$  & $\gamma_1$  & $\gamma_2$  & $\gamma_3$  & $\gamma_4$  & $\gamma_5$  & P (\%) & R (\%) & F (\%) & $\gamma_i$    & P (\%) & R (\%) & F (\%) 
& P (\%) & R (\%) & F (\%) \\ \hline
0th \& 1st         & 0.3 & 1   & x   & x   & x   & x   & 77.72   & 83.51 & 80.51 & 0.33 & 78.05   & 83.57 & 80.71 
& 77.67 & 78.43 & 78.05\\
1st \& 2nd         & 0.3 & 0.5 & 1   & x   & x   & x   & 80.55   & 87.20 & 83.74 & 0.48 & 81.05   & 85.23 & 83.09 
& 79.35 & 88.10 & 83.50\\
2nd \& 3rd         & 0.2 & 0.6 & 0.9 & 1   & x   & x   & 79.33   & 89.93 & 84.30 & 0.71 & 80.28   & 86.66 & 83.35 
& \textbf{82.61} & 85.99 & \textbf{84.26}\\
3rd \& 4th         & 0.1 & 0.9 & 0.9 & 0.5 & 1   & x   & 81.06   & 90.80 & 85.66 & 0.65 & 80.16   & 87.08 & 83.47 
&81.02 & 85.83 & 83.36\\
4th \& 5th         & 0.1 & 0.9 & 0.9 & 0.7 & 0.8 & 1   & 81.14   & 90.67 & 85.64 & 0.78 & 80.20   & 87.27 & 83.58 
& 81.80 & \textbf{86.22} & 83.95\\
5th \& 6th         & 0.1 & 0.9 & 0.9 & 0.7 & 0.8 & 0.5 & \textbf{82.35}   & \textbf{91.10} & \textbf{86.50} & 0.70 & \textbf{80.29}   & \textbf{87.57} & \textbf{83.77} 
& 81.97 & 82.18 & 82.07\\ \hline
\end{tabular}
\end{table*}

We apply the MLC-CFP method for MF0 estimation of polyphonic music data, and investigate the effects of 1) the number of layers, 2) critical parameters $\gamma_i$ of each layer, and 3) the presence of convolutional noise. 
We used the Bach10 dataset, containing ten quartets of four different instruments 
\cite{duan2010multiple} 
for the evaluation of multi-pitch estimation (MPE). 
To see the behavior of the algorithm in a straightforward way, we consider three methods to optimize parameters $\{\gamma_i\}^{L}_{i=0}$ for an $L$-layer MLC-CFP algorithm combining $\vZ^{(L-1)}$ and $\vZ^{(L)}$: 

1) the \emph{brute-force method} that searches for the optimal parameters exhaustively for each $\gamma_i$. The search range of every $\gamma_i$ is from 0.1 to 0.9 with step size of 0.1. 

2) the \emph{greedy method} that searches for the optimal value of $\gamma_l$ based on the optimal values of $\{\gamma_i\}^{l-1}_{i=0}$. The search range is from 0.01 to 0.99 with step size of 0.01, since its light computation allows us to use finer grids. For the post-processing of the brute-force and the greedy methods, we use the CFP-based algorithm reported in \cite{su2017HSP_DNN}, as a simplified version of \cite{su2015combining} without too detailed hand-crafted rules.

3) the \emph{stochastic gradient descent (SGD) method} that minimizes the binary cross-entropy between $\vY^{(L)}$ (with 88 log-frequency filterbanks and a sigmoid function as the output layer) and the ground-truth piano roll by performing SGD over every $\gamma_i$. 
We apply non-repeated 10-fold cross-validation on the dataset: for each time one music piece is chosen as the test data and others as training data. The resulting F-scores are obtained by 
summing all the counts of the true positives, false positives, and false negatives. We initialize $\gamma_0$ to 0.24, $\gamma_1$ to 0.6, and all other $\gamma_i$s to 1, and use the SGD optimizer with 0.1 learning rate, a mini-batch size of 256, and train the network up to 40 epochs. 

For other parameters, we compute 7939-point DFT using Blackman-Harris window, the hop size is 10 ms, the cutoff frequency $f_c$ is 27.5 Hz (frequency of \texttt{A0}), and the cutoff quefrency $q_c$ is 0.24 ms (period of \texttt{C8}). 
Since there are many task-dependent parameters, all the source codes will be announce after this paper is accepted for reproducibility.

Table \ref{table:comparison} lists the optimal precision (P), recall (R), F-score (F), along with $\gamma_i$ for every pairs $\left[\vZ^{(L-1)}, \vZ^{(L)}\right]$ for all optimization approaches. 
When using the brute-force method, the resulting P, R and F all increase consistently when the number of layers increases. 
Note that the recall increases dramatically with higher-order layer because the weak F0 peaks are enhanced, similar to the simulation example in Fig. \ref{fig:2layer_ddft}.  
When using the greedy method, we observe that although the performance is sub-optimal, it still increases as $L$ increases. 
When using gradient descent, there is no consistent improvement in F-scores related to the number of layers, possibly because the hand-crafted rules such as harmonic selection \cite{su2015combining} could not be applied to this model. 
However, it still gives competitive performance for $L=3$. More specifically, when compared to other MF0 estimation methods, all the three methods in our discussion outperform the state-of-the-art matrix factorization approaches such as constrained NMF and probabilistic latent component analysis (PLCA) (see Table III in \cite{su2015combining} for more details). The brute-force method also outperform the CFP result reported in \cite{su2015combining}, in which the parameters for the hand-crafted rules are also based on brute-force searching. 



\begin{figure}[t]
\begin{tikzpicture}[domain=10:1000]
    \begin{semilogxaxis}[
    legend pos=south west,
    xlabel = {Cutoff frequency (Hz)},
    ylabel = {F-score}]
        \addplot+[mark=*] coordinates {
        ( 10, 0.8096 )
        ( 25, 0.8092 )
        ( 50, 0.8084 )
        ( 100, 0.8072 )
        ( 250, 0.7993 )
        ( 500, 0.6736 )
        ( 1000, 0.4718 )
        };
        \addplot+[mark=o] coordinates {
        ( 10, 0.8350 )
        ( 25, 0.8332 )
        ( 50, 0.8324 )
        ( 100, 0.8321 )
        ( 250, 0.7958 )
        ( 500, 0.7011 )
        ( 1000, 0.6207 )
        };
        \addplot+[mark=asterisk] coordinates {
        ( 10, 0.8428 )
        ( 25, 0.8455 )
        ( 50, 0.8465 )
        ( 100, 0.8484 )
        ( 250, 0.8257 )
        ( 500, 0.7452 )
        ( 1000, 0.6675 )
        };
        \addplot+[dotted, mark=triangle*] coordinates {
        ( 10, 0.8574 )
        ( 25, 0.8589 )
        ( 50, 0.8624 )
        ( 100, 0.8664 )
        ( 250, 0.8638 )
        ( 500, 0.8006 )
        ( 1000, 0.7089 )
        };
        \addplot+[mark=otimes] coordinates {
        ( 10, 0.8565 )
        ( 25, 0.8566 )
        ( 50, 0.8600 )
        ( 100, 0.8634 )
        ( 250, 0.8657 )
        ( 500, 0.8156 )
        ( 1000, 0.7295 )
        };
        \addplot+[mark=square*] coordinates {
        ( 10, 0.8642 )
        ( 25, 0.8648 )
        ( 50, 0.8688 )
        ( 100, 0.8710 )
        ( 250, 0.8707 )
        ( 500, 0.8140 )
        ( 1000, 0.7274 )
        };
        \legend{$\vZ^{(0)}$ \& $\vZ^{(1)}$,$\vZ^{(1)}$ \& $\vZ^{(2)}$,
        $\vZ^{(2)}$ \& $\vZ^{(3)}$,$\vZ^{(3)}$ \& $\vZ^{(4)}$,
        $\vZ^{(4)}$ \& $\vZ^{(5)}$,$\vZ^{(5)}$ \& $\vZ^{(6)}$}
	\end{semilogxaxis}
    
\end{tikzpicture}
\caption{F-scores of Bach10 dataset under high-pass noise.}
\label{fig: highpass}
\end{figure}
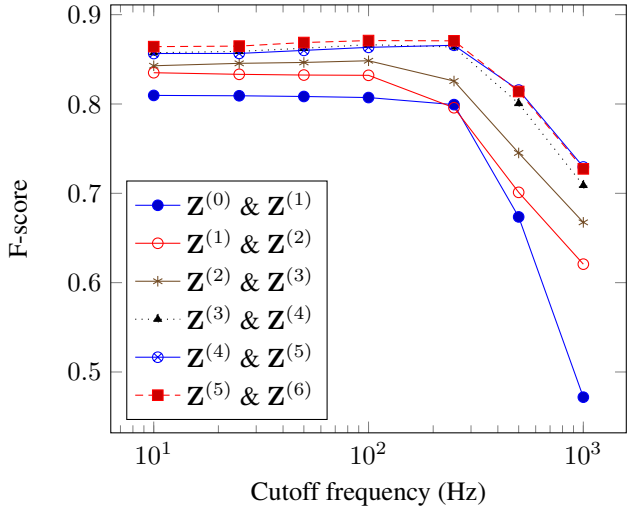

Fig. \ref{fig: highpass} shows the resulting F-scores on the Bach10 dataset with high-pass degradation, which is simulated by a 4th-order Butterworth high-pass filters with cutoff frequency from 10 to 1000 Hz. The parameters used here are those found from the brute force method; see Table \ref{table:comparison}. Fig. \ref{fig: highpass} clearly shows that with more layers the results are better: even under a severe cutoff frequency at 1 kHz, a 6-layer MLC still gives 72.74\% F-score, 25\% higher than the one of a 1-layer MLC. Even more interestingly, results in some degradation conditions are better than those in the normal condition. For example, pairs $\vZ^{(5)}$ and $\vZ^{(6)}$ achieve 87.10\% F-score with cutoff frequency at 100 Hz; this is probably because the simulated degradation suppresses the low-frequency noise incidentally.

To summarize, our experiments show that 1) in the MLC-CFP algorithm, a deep DFT do performs better than a shallow DFT, and 2) MLC is highly robust to high-pass noise.

\section{Discussion and conclusion}

We have introduced the MLC, explained its physical insight in signal processing, and verified its potential in detecting multiple F0s through simulation and real-world data under normal and noisy conditions. 
We also emphasize that the MLC is not only useful in the signals contaminated with high-pass convolutional noise, but also useful in all signals which inherently have weak F0 magnitude. For example, the F0 of bassoon in the Bach10 dataset is usually weaker than other instruments (violin, clarinet and saxophone), but experiments show that this issue is addressed by the MLC. 
From the result in the filter test we also suggest that in practice using MLC with a little high-pass filter can get better result.


One risk of MLC is that it may generate fake F0 peaks. For example, two F0s at $a$ Hz and $b$ Hz may produces a cross term at $|a-b|$ Hz due to effect of nonlinear activation. This cross term cannot be suppressed and may be even enhanced because it appears in both the time-domain and frequency-domain representations. To restrict the strength of cross terms, 
the results indicate a possible strategy: setting a small $\gamma_0$ and larger $\gamma_i$s at the higher layers. 
Our pilot study also shows that even fake cross terms are produced, they are mostly weak unless all $\gamma_i$s approach zero.

Developing a more efficient way to optimize the parameters efficiently on a large-scale dataset with complicated sources and harmonic structures will be another focus in the future. From the current finding, we summarize that MLC-CFP works well in general and is relatively easy to optimize for $L\leq 3$, but a great potential exists in a deeper MLC.



\bibliographystyle{IEEEbib}
\bibliography{DDFT}

\end{document}